%% file: jwsmith_TOP18.tex
\documentclass[11pt]{article}
\usepackage{graphicx}
\usepackage{booktabs}
\usepackage{subfig}
\usepackage{xspace}
\usepackage{amsmath}
\usepackage{multirow}
\usepackage{amssymb}
\usepackage{url}
 \usepackage{cite}
\usepackage[
    letterpaper,
    bindingoffset=-0.1in,
    centering
    ]{geometry} 
    
\usepackage{lineno}
\usepackage{placeins}
\newlength{\bibitemsep}
\newlength{\bibparskip}
\let\oldthebibliography\thebibliography
\renewcommand\thebibliography[1]{%
  \oldthebibliography{#1}%
  \setlength{\parskip}{\bibitemsep}%
  \setlength{\itemsep}{\bibparskip}%
}
\def\TeV{\ifmmode {\mathrm{\ Te\kern -0.1em V}}\else
                   \textrm{Te\kern -0.1em V}\fi\xspace}%
\def\GeV{\ifmmode {\mathrm{\ Ge\kern -0.1em V}}\else
                   \textrm{Ge\kern -0.1em V}\fi\xspace}%
\def\MeV{\ifmmode {\mathrm{\ Me\kern -0.1em V}}\else
                   \textrm{Me\kern -0.1em V}\fi\xspace}%
\def\keV{\ifmmode {\mathrm{\ ke\kern -0.1em V}}\else
                   \textrm{ke\kern -0.1em V}\fi\xspace}%
\def\eV{\ifmmode  {\mathrm{\ e\kern -0.1em V}}\else
                   \textrm{e\kern -0.1em V}\fi\xspace}%

\def\pt{\ensuremath{p_{\mathrm{T}}}\xspace} 

\def\MET{\ensuremath{E_{\mathrm{T}}^{\mathrm{miss}}}\xspace} 
\newcommand*{\mwt}{$m_{\text{T}}(W)$\xspace}

\newcommand*{\xsecModifyNoun}{cross-section\xspace}
\newcommand*{\ttgamma}{$t\bar{t}\gamma$\xspace}

\newcommand*{\Wgamma}{$W\gamma$\xspace}
\newcommand*{\Zgamma}{$Z\gamma$\xspace}

\newcommand*{\Other}{Other prompt\xspace}

\newcommand*{\ttV}{$t\bar{t}V$\xspace}
\newcommand*{\ttbar}{$t\bar{t}$\xspace}
\newcommand*{\hfake}{hadronic fake\xspace}

\newcommand*{\efake}{$e\to \gamma$ fake\xspace}

\newcommand*{\chljets}{single-lepton\xspace}
\newcommand*{\chll}{dilepton\xspace}
\newcommand*{\chejets}{$e$+jets\xspace}
\newcommand*{\chmujets}{$\mu$+jets\xspace}
\newcommand*{\chee}{$ee$\xspace}
\newcommand*{\chemu}{$e\mu$\xspace}
\newcommand*{\chmumu}{$\mu\mu$\xspace}

\def\institute{II. Physikalisches Institut\\
Goerg-August Universit\"at, Germany}

\def\Title#1{\begin{center} {\Large #1 } \end{center}}
\def\Author#1{\begin{center}{ \sc #1} \end{center}}
\def\Address#1{\begin{center}{ \it #1} \end{center}}

\newenvironment{Abstract}{\begin{quotation}  }{\end{quotation}}
\newenvironment{Presented}{\begin{quotation} \begin{center} 
             PRESENTED AT\end{center}\bigskip 
      \begin{center}\begin{large}}{\end{large}\end{center} \end{quotation}}

\input econfmacros.tex

\begin{document}

\begin{titlepage}
\vfill
\Title{Fiducial cross-section measurements of top-quark pair production in association with a photon at $\sqrt{s}$ = 13 TeV with the ATLAS
detector}
\vfill
\Author{ Joshua Wyatt Smith}
\Address{on behalf of the ATLAS collaboration \vspace{0.5cm} \\ \institute}
\vfill
\begin{Abstract}
\noindent Top-quark pairs in association with final state particles are produced in large quantities at the LHC due to the high centre-of-mass energy available in proton-proton collisions. One such topology is that of a prompt photon radiated from the top quark in addition to the final state particles from the top quark decay. 
Using 36.1 fb$^{-1}$ of data collected at $\sqrt{s}$ = 13 TeV with the ATLAS detector, fiducial cross-section results are shown in the single-lepton and dilepton channels. Object-level and event-level neural networks are used to increase sensitivity.
\end{Abstract}
\vfill
\begin{Presented}
$11^\mathrm{th}$ International Workshop on Top Quark Physics\\
Bad Neuenahr, Germany, September 16--21, 2018
\vspace{2cm}

\small{Copyright 2019 CERN for the benefit of the ATLAS Collaboration. Reproduction of this article or parts of it is allowed as specified in the CC-BY-4.0 license.}
\end{Presented}
\vfill
\end{titlepage}
\def\thefootnote{\fnsymbol{footnote}}
\setcounter{footnote}{0}

\section{Introduction}
The production of a top-quark pair in association with a photon is a direct probe of the electromagnetic coupling of the top quark.
Possible anomalous top-quark couplings could manifest as shape discrepancies in various kinematic distributions or in cross-section measurements~\cite{Baur:2004uw,Bouzas:2012av}. The results can also be interpreted in the framework of an effective field theory in the search for new physics~\cite{eft_1601.08193}.
Photons can be radiated from all charged particles, including those in top quark decay and incoming initial state quarks. Kinematic cuts are applied to reduce the contribution of photons being radiated from top quark decay.

Previous studies have measured the cross section in the single-lepton channels~\cite{Aaltonen:2011sp,Aad:2015uwa,Sirunyan:2017iyh,TOPQ-2015-21}. 
In the analysis~\cite{ATLAS-CONF-2018-048} summarised here, the fiducial cross sections are measured in the single-lepton channels and, for the first time, the dilepton channels. No distinction is made between electrons and muons from tau lepton decays.
The measurements are performed using the ATLAS~\cite{PERF-2007-01} detector at the LHC and a total dataset corresponding to 36.1~fb$^{-1}$, collected in 2015 and 2016 at a centre-of-mass energy of 13~TeV. 
Dedicated $k$-factors applied to the leading-order signal sample allow to compare the extracted cross sections to the next-to-leading order (NLO) theoretical calculations~\cite{melnikov}. 

\section{Signal region definition}
Event selections are made to enhance the purity of the \ttgamma signal.
Electrons are required to have $\pt > 25$~GeV ($\pt > 27.5$~GeV) for the 2015 (2016) data, while muons are required to have $\pt > 27.5$~GeV for the 2015 and 2016 data. Only one reconstructed lepton is required for the single-lepton channels, while exactly two leptons of opposite sign are required for the dilepton channels. Electrons and muons are both required to be isolated based on calorimeter and track information.
At least four (two) jets are required in the \chejets and \chmujets (\chee, \chemu and \chmumu) channels. At least one of these reconstructed jets should be tagged as a $b$-jet with a working point efficiency of 77\%.
In all the \chll channels a minimum invariant mass between the two leptons is such that $m(l,l) > 15$~\GeV. 
In the \chee and \chmumu channels values of the invariant mass of the two leptons close to the mass of the $Z$-boson are vetoed, such that $m(l,l) \notin$ [85,95]~\GeV. The same requirement is applied to the invariant mass of the system of the lepton pair and photon, $m(l,l,\gamma)$.
In the \chee and \chmumu channels a cut on the missing transverse energy of an event is required to be above a threshold, $\MET > 30$~\GeV.
In the \chejets channel a veto is placed on the invariant mass of the photon and the electron such that $m(\gamma,e) \notin$ [85,95]~\GeV. 
In all channels, exactly one photon of $\pt>20$~\GeV is required. This photon needs to be isolated.
A final cut is applied on the $\Delta R$ distance between the photon and the leptons, $\Delta R(\gamma,l) > 1.0$. This cut reduces the contribution of photons from top quark decay products.

The fiducial phase space is defined at particle level such that it mimics the above event selection, with the exception of the invariant mass veto cuts and the \MET cuts that are only intended to suppress the background.

\section{Background processes}
The main background in the single-lepton channels consists of photons which are misidentified as electrons, the so-called ``\efake{}" background.
This occurs mainly from the \ttbar dileptonic decays, specifically from the \chemu and \chee channels. This background is negligible in the \chll channels.
Scale factors derived from a data-driven \emph{tag-and-probe} method are used to correct the number of fake photons predicted by the Monte Carlo (MC) samples.

Photons from hadrons, or hadrons misidentified as photons make up the ``\hfake{}" background.
The majority of \hfake photons comes from \ttbar events. Small contributions arise from $W/Z$+jets and single-top quark processes.
Scale factors to correct MC prediction to data are derived in the single-lepton channel using the ``ABCD" method. These are extrapolated to the dilepton channels.

Events with non-prompt leptons as well as fake leptons may satisfy the event selections for the signal region.
These are categorised as ``fake lepton" background. The main contribution comes from the QCD driven multi-jet processes in association with a photon.
The estimation of the fake lepton background in the single-lepton channels follows the fully data-driven Matrix Method approach. In the dilepton channels this contribution is negligible.

Events in which a real prompt photon is radiated from anything except top quarks (or incoming quarks) make up the ``\Other{}" background.
For the \chljets channel this is largely due to the \Wgamma process, whereas for the \chll channel the main contribution is from the \Zgamma process. Other processes such as single-top, diboson and \ttV{}, where each has an additional prompt photon in the final state, have small or even negligible contributions. 
In the \chljets (\chll) channels the \Wgamma (\Zgamma) contribution is largest and is therefore separated from the rest of the prompt photon backgrounds. The ``\Other{}" background is estimated using MC.

\section{Analysis Strategy}

Object-level and event-level neural networks (NNs) are used to improve the sensitivity of the measurement in the single-lepton channels, while only an event-level NN is used in the dilepton channels. 
All NNs are implemented based on a feedforward architecture in the Keras~\cite{chollet2015keras} framework using the Theano~\cite{2016arXiv160502688short} backend. 
The \textsc{lwtnn}~\cite{lwtnn} library is used to inject the trained NNs back into the ATLAS software framework.
The object-level NN, called the Prompt Photon Tagger (PPT), is trained on independent prompt photon and di-jet samples, using information from the calorimeters. This NN serves to distinguish between prompt photons and hadronic fake photons. 
In the case of the \chljets channel, the PPT distribution is then used as an input into the event-level NN, called the Event-Level Discriminator (ELD). 
The \chljets (\chll) ELD is trained using 15 (7) variables. For the \chljets ELD, information about the $b$-tagged jet and the PPT proves to provide the most powerful variables. For the \chll ELD, the $b$-tagged jet and the invariant mass of the two leptons prove to be very powerful. Examples of other variables used in the training include the \MET, \pt of various jets and \mwt (for the \chljets channel).

The fiducial cross sections for each channel are extracted by performing a maximum likelihood fit to the ELD distribution. The cross section ($\sigma_{\text{fid}}$) is related to the number of signal events by 
\begin{equation}
N^{s}_i = \mathcal{L} \times \sigma_{\text{fid}} \times C \times f_{i}^{\text{ELD}},
\end{equation}
where $\mathcal{L}$ is the integrated luminosity, $C$ is the ratio of all simulated events passing event selections to those only falling into the fiducial phase space, and  $f_{i}^{\text{ELD}}$ is the fraction of events falling into bin $i$ of the ELD.

\section{Results}

\begin{figure}[!htbp]
\centering
\subfloat[]{
\includegraphics[width=0.42\linewidth]{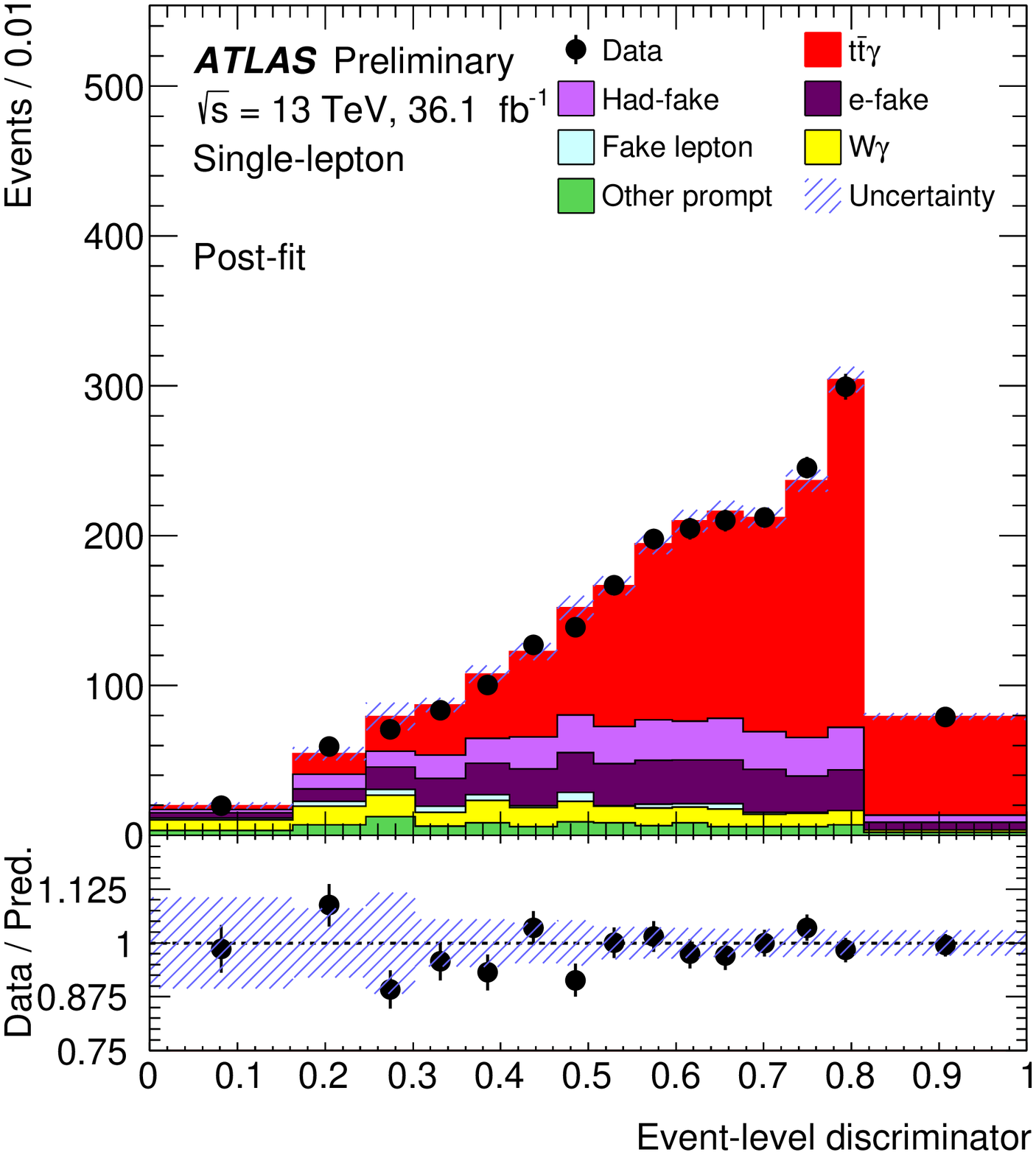}
}
\subfloat[]{
\includegraphics[width=0.42\linewidth]{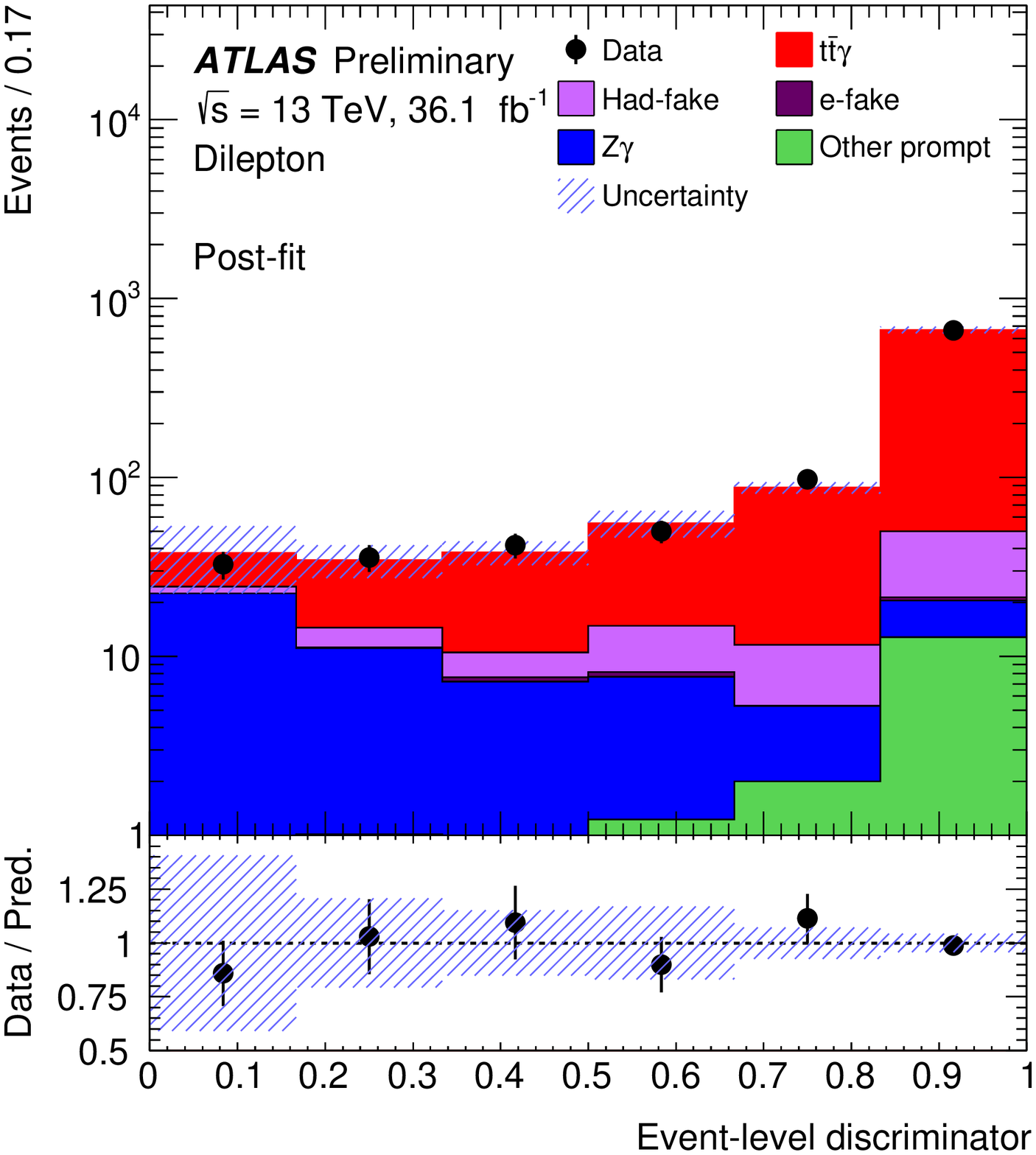}
}
\caption [] {The post-fit ELD distributions for the (a) single-lepton and (b) dilepton channels.
All the systematic uncertainties are included~\cite{ATLAS-CONF-2018-048}.}
\label{fig:ELDpostfit}
\end{figure}

Eight cross-section measurements are extracted from a maximum likelihood fit in the \chljets, \chll and combined channels. The post-fit ELD distributions for two of the measurements (the \chljets and \chll channel) are shown in Figure~\ref{fig:ELDpostfit}, where good agreement can be seen between MC and data points. The ELD distributions for other channels look similar.
Table~\ref{tab:systematics} shows the effects that groups of systematic uncertainties have on the fiducial cross-section measurements. In the \chljets channel, the largest contributions come from the modelling of jets, the backgrounds, and the PPT. In the \chll channel, signal and background modelling have the largest effect, but are generally still small.

The eight extracted cross sections are summarised in Table~\ref{tab:fidcrosSections}.
In summary, all measurements are in agreement with NLO predictions.

\begin{table}[!ht]
\centering 
\scalebox{0.8}{
\begin{tabular}{l|r|r}
\toprule
Source &  Single lepton (\%) &  Dilepton (\%) \\
\hline
Signal modelling 	& $\pm$ 1.6 	& $\pm$ 2.9   \\
Background modelling	& $\pm$ 4.8 	& $\pm$ 2.9   \\
Photon			& $\pm$ 1.1 	& $\pm$ 1.1   \\
Prompt-photon tagger 	& $\pm$ 4.0 	&   -    \\
Leptons			& $\pm$ 0.3 	& $\pm$ 1.3   \\
Jets			& $\pm$ 5.4 	& $\pm$ 2.0   \\
$b$-tagging		& $\pm$ 0.9 	& $\pm$ 0.4   \\
Pile-up			& $\pm$ 2.0 	& $\pm$ 2.3   \\
Luminosity		& $\pm$ 2.3 	& $\pm$ 2.3   \\
MC sample size		& $\pm$ 1.9 	& $\pm$ 1.7   \\
\hline
Total systematic uncertainty & $\pm$ 7.9 & $\pm$ 5.8 \\
Data sample size   & $\pm$ 1.5 & $\pm$ 3.8 \\
\hline
Total uncertainty & $\pm$ 8.1 & $\pm$ 7.0 \\
\bottomrule
\end{tabular}
}
\caption{
    Summary of the effects of the groups of systematic uncertainties on the fiducial cross section in the \chljets and \chll channels~\cite{ATLAS-CONF-2018-048}.}
\label{tab:systematics} 
\end{table}

\begin{table}[!ht]
\centering 
\scalebox{0.8}{
\begin{tabular}{|l|c|c|}
 \hline
  & \multicolumn{2}{c|}{$\sigma^{\text{fid}}_{\text{\ttgamma}}$ [fb]} \\ \hline
Channel & Measured $\pm$(stat.) $\pm$(syst.) & Theory $\pm$(total) \\ \hline
\chejets & $265 \pm 6 \pm 21$   &  $247 \pm 49$ \\ 
\chmujets & $250 \pm 7 \pm 22 $   &  $248 \pm 50$ \\ 
\chee & $16 \pm 2 \pm 2 $   &  $16 \pm 2$ \\ 
\chmumu & $18 \pm 1 \pm 2 $   &  $16 \pm 2$ \\ 
\chemu & $34 \pm 2 \pm 2 $   &  $31 \pm 5$ \\ \hline

Single-lepton & $521 \pm 9 \pm 41 $   &  $495 \pm 99$ \\ 
Dilepton & $69 \pm 3 \pm 4 $   &  $63 \pm 9$ \\ \hline

Inclusive (5 channels) & $589 \pm 10 \pm 34 $   &  $558 \pm 110$ \\ 
\hline
\end{tabular}
}
\caption{Fiducial \xsecModifyNoun summary for all channels, as well as theoretical predictions~\cite{ATLAS-CONF-2018-048}.}
\label{tab:fidcrosSections} 
\end{table}

\FloatBarrier


\bibliographystyle{atlas}
\bibliography{Top18}

\end{document}

%% file: econfmacros.tex
\def\beq{\begin{equation}}
\def\eeq#1{\label{#1}\end{equation}}
\def\eeqn{\end{equation}}

\def\beqa{\begin{eqnarray}}
\def\eeqa#1{\label{#1}\end{eqnarray}}
\def\eeqan{\end{eqnarray}}

\let\bar=\overbar

\def\Dslash{\not{\hbox{\kern-4pt $D$}}}
\def\dslash{\not{\hbox{\kern-2pt $\del$}}}

\def\msb{{\bar{\ssstyle M \kern -1pt S}}}

